\documentclass[11pt]{article}
\usepackage{a4wide}
\usepackage{authblk}
\usepackage{graphicx,xcolor}
\usepackage{caption}
\usepackage{subcaption}
\usepackage{float}
\usepackage{color}
\usepackage{amsmath,amsfonts,amssymb,graphicx,hyperref,hypcap}
\usepackage{epsfig}
\usepackage{amsmath,amsfonts,epsf}
\usepackage{amssymb,tikz}

\begin{document}
\begin{center}
{\Large\bf  Time-dependent diffusive interactions of dark matter and dark energy
 in the context of $k-$essence cosmology}
\end{center}
\vspace{4mm}
\begin{center}
{\large Abhijit Bandyopadhyay\footnote{Email: abhijit@rkmvu.ac.in} and Anirban Chatterjee\footnote{Email: anirban.chatterjee@rkmvu.ac.in}}
\end{center}

\begin{center}
Department of Physics\\Ramakrishna Mission Vivekananda Educational and Research Institute\\
Belur Math, Howrah 711202, India
\end{center}
\vspace{4mm}

\abstract{We investigated the scenario of 
time-dependent diffusive interaction 
between dark matter and dark energy
and showed that such a model can be accommodated
within the observations of luminosity distance - redshift data  in Supernova Ia
(SNe Ia) observations.
We obtain   constrains on  different relevant parameters of this model from the
observational data. We consider a homogeneous scalar field $\phi(t)$ driven by
a $k-$essence Lagrangian of the from $L = V(\phi)F(X)$ with constant 
potential $V(\phi) = V$, to describe the dynamics of dark energy in this model.
Using the temporal behaviour of the FRW scale factor, the equation of state
and total energy density of the dark fluid, extracted from the analysis 
of  SNe Ia (JLA) data, we have obtained the time-dependence of the $k-$essence
scalar field and also reconstructed  form of the function $F(X)$
in the $k-$essence Lagrangian.
}

\section{Introduction}           
\label{sec:intro}
In this work, we explore a scenario of interacting dark matter (DM) and dark energy (DE), where the DE is represented by a homogeneous scalar field $\phi$ with its dynamics driven by a $k$-essence Lagrangian with constant potential and DM is assumed to be a perfect fluid.
In the late time phase of cosmic evolution, we can neglect the effect of radiation and baryonic matter to the total energy density of the universe. Thus for the late time phase, dark energy and dark matter are the dominant sectors to study the new consequences. Motivations behind consideration of such classes of a unified model of DM and DE is to address the issue of coincidence of observed present-day dark energy and dark matter densities using a dynamical relation between dark matter and dark energy \cite{DM-DE,Bandyopadhyay1}. The interaction between DM and DE is assumed to happen through velocity diffusion of particles of dark matter fluid 
in the background   of field $\phi$.  In an earlier work \cite{Bandyopadhyay2},
we have studied implications of such a diffusive interaction with 
constant (time independent) diffusion coefficient.
In this work we revisit interacting model of DE and DM
with a time-dependent diffusion coefficient. We also explore the 
implications of  measurements of luminosity distances and redshifts 
in Supernova Ia (SNe Ia) observations \cite{ref:Suzuki,Betoule:2014frx,Wang1,Wang} 
in the constraining different parameters for this model.\\

The basic framework of the model of DM-DE interactions 
has been discussed in detail in
\cite{Bandyopadhyay2,Velten}. We, first, indulge in a brief recollection of the model
with an emphasis on the consideration of the time-dependence of the
diffusion coefficient.
We assume dark matter and dark energy to be perfect fluids
and neglect contribution of radiation and baryonic matter
during the late time phase of cosmic evolution (probed
in SNe Ia observations).  In a flat FRW spacetime background,
conservation of total energy momentum tensor for the 
dark fluid (DE + DM) takes the form
\begin{eqnarray}
\Big{[} \dot{\rho}_{\rm dm} + 3H\dot{\rho}_{\rm dm}\Big{]}
= - \Big{[} \dot{\rho}_{\rm de} + 3H ( \rho_{\rm de} +  p_{\rm de})\Big{]}
= Q(t)
\label{eq:cont1}
\end{eqnarray}
where $\rho_{\rm dm}$ and $\rho_{\rm de}$
are the energy densities of dark matter and dark energy fluid 
respectively. $p_{\rm de}$ is the pressure of the
dark energy fluid and we take dark matter as
pressureless dust. $H \equiv \dot{a}/a$ is the Hubble parameter,
where $a(t)$ is the scale factor
corresponding to Friedman-Robertson-Walker (FRW) spacetime background. 
Here $t$ is regarded as a dimensionless time parameter 
with $t=1$ corresponding to present epoch.
The quantity $Q(t)$ in Eq.\  (\ref{eq:cont1}),
is a measure of  rate of energy transfer between the fluid
dark matter and dark energy caused by a diffusion 
at the instant of time $t$.
We parametrize time dependence of the quantity $Q(t)$ in terms of
a parameter $k$  
\begin{eqnarray}
Q(t)=Q_0  \big{[}a(t) \big{]} ^k
\label{eq:qt}
\end{eqnarray}
where scale factor $a(t)$ has been taken to be normalised to unity
at present epoch, $a(t=1)=1$,  and $Q_0$ is the value of $Q(t)$ at present epoch.
Using the results of a comprehensive analysis of 
recently released ``Joint Light-curve Analysis" (JLA) data 
\cite{Betoule:2014frx} consisting of 740 SNe Ia events
as performed in \cite{Wang1}, the observed temporal behaviour 
of the quantities: the scale factor $a(t)$, the total energy density $(\rho_{\rm dm}  + \rho_{\rm de})(t)$
and the equation of state $\omega(t)$ have been extracted over 
a time domain $0.44 < t < 1$ accessible in SNe Ia observations.
Using these observed time evolutions and the  chosen form Eq.\  (\ref{eq:qt})
of parametrisation of   $Q(t)$   in Eq.\  (\ref{eq:cont1})
we  obtained the range of values of $k$ for which
the interacting DE-DM  model with such a time-dependent 
diffusion term $Q(t)$, may be accommodated within the
scheme of luminosity distance - redshift measurements of
SNe Ia observation.  The methodology of obtaining such constraints
has been discussed comprehensively in Sec.\ \ref{sec:observation}.     \\

We have assumed that, the diffusive interaction, considered here, 
occurs in the background of $k-$essence scalar field $\phi$, 
whose dynamics is driven by a   
non-canonical Lagrangian $L=V(\phi)F(X)$, where $X=(1/2)g^{\mu\nu}\nabla_\mu\phi \nabla_\nu\phi$.
We take the potential $V(\phi)$ to be constant, which ensures existence of a scaling relation 
of the form $X(dF/dX)^2=Ca^{-6}$, where $C$ is a constant \cite{scale1,scale2}. 
We then identify the stress energy tensor corresponding to the 
non-canonical Lagrangian   with that of the
dark energy fluid and consider the $k-$essence scalar field $\phi = \phi(t)$ to
be  homogeneous. Use of the scaling relation, then enables us to establish
relations between  time derivative of scalar field  and  energy density and pressure of dark energy fluid.
Using the above relations along with the observed  temporal behaviour of the quantities like
$a(t)$, $\omega(t)$ and $\rho_{\rm dm} + \rho_{\rm de}$ we reconstructed  the temporal behaviour of the $k-$essence scalar field $\phi$
as well as the form of the function $F(X)$. The results are shown for
different benchmark values of the parameter $k$ with its allowed domain as obtained earlier.   The quantity $Q(t)$ plays an important role to study the interaction between DM and DE sectors. We have also shown the energy transferred in these two sectors for some chosen values of the diffusion parameters. This has been described in detail in Sec.\ \ref{sec:k-essence}.

\section{Observational constraints on time-dependent diffusive DE-DM interacting model}
\label{sec:observation}
\begin{figure}[t]
\begin{center}
\includegraphics[scale=0.35]{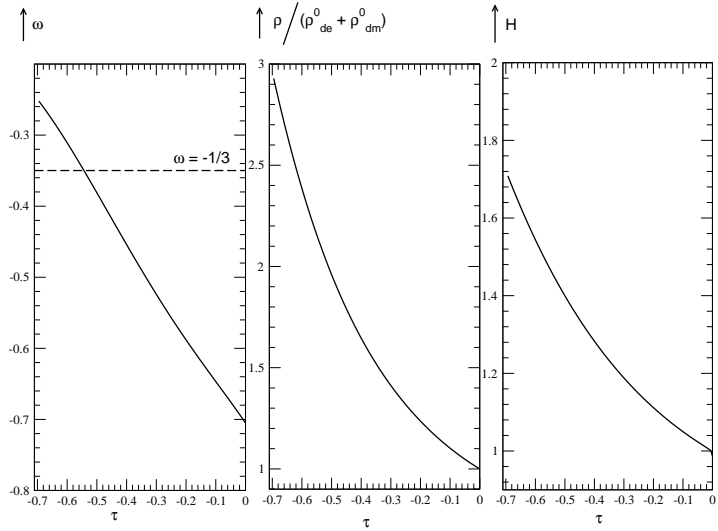} 
\end{center}
\caption{\label{fig:1}
Plot of   $\omega$, $\rho / (\rho^0_{\rm de} + \rho^0_{\rm dm})$ and $H$ as a function 
of time parameter $\tau$ as extracted from JLA data.}
\end{figure}

We use the measurements of luminosity distance and redshift 
in SNe Ia observations for redshift values up to $z \approx 1$ to extract features of 
late time phase of cosmic evolution. These data are instrumental
in obtaining the temporal behaviour of FRW scale factor $a(t)$.
There exist different compilations of SNe Ia data from different
surveys. These include
high redshift ($z \sim 1$) projects \textit{viz.} 
Supernova Legacy Survey (SNLS) 
(\cite{ref:Astier},\cite{ref:Sullivan}), the ESSENCE project 
\cite{ref:Wood-Vasey}, 
the Pan-STARRS survey (\cite{ref:Tonry},\cite{ref:Scolnic},
\cite{ref:Rest}); intermediate redshift  ($0.05<z<0.4$)
projects \textit{viz} SDSS-II supernova surveys (\cite{ref:Frieman},\cite{ref:Kessler},\cite{ref:Sollerman}, \cite{ref:Lampeitl},\cite{ref:Campbell}) and small redshift programmes such as
surveys like   Harvard-Smithsonian Center for Astrophysics survey   
\cite{ref:Hicken}, 
the Carnegie Supernova Project  (\cite{ref:Contreras},\cite{ref:Folatelli},\cite{ref:Stritzinger}),
the Lick Observatory Supernova Search  
\cite{ref:Ganeshalingam}  and the Nearby Supernova Factory  \cite{ref:Aldering}. Various other compilations of SNe Ia data may 
be found in  \cite{snrest}. All such surveys include nearly one thousand
SNe Ia events. In this work we used the results from analysis
of ``Joint Light-curve Analysis" (JLA) data (\cite{ref:Suzuki},\cite{Betoule:2014frx},\cite{Wang1},\cite{Wang}) consisting of   
 740 data points from  3 years survey of SDSS,  five-year SNLS survey and 14 very
high redshift $0.7 < z < 1.4$ SNe Ia from space-based observations with the HST\cite{ref:Riess}.
The methodology of analysis of JLA data has been described in detail
in \cite{Wang1,Bandyopadhyay3} where a $\chi^2$ function for the JLA data 
 is defined as 
\begin{eqnarray}
\chi^2_{\rm SN} (\alpha,\beta) &=& \sum_{i,j}\left(\mu_{{\rm obs}}^{(i)} - \mu_{\rm th}^{(i)}\right)
 \left(\Sigma^{-1}\right)_{ij}  \left(\mu_{\rm obs}^{(j)} - \mu_{\rm th}^{(j)}\right) 
\label{eq:chisqsn}
\end{eqnarray}
where $i$ and $j$ run from 1 to 740 corresponding to   740 SNe Ia events  
contained in the JLA data set \cite{Betoule:2014frx}. $\mu_{{\rm obs}}^{(i)}$ is the observed value of distance modulus at a redshift $z_i$ corresponding to $i^{\rm th}$ entry of the JLA data set and 
$\mu_{{\rm th}}^{(i)}$ is the corresponding theoretical estimate
expressed through an empirical relation expressed in terms of various 
parameters. $\Sigma$ is the total covariant matrix given in terms of statistical and systematic uncertainties
(see \cite{Wang1} for details). In \cite{Wang1}, Wang {\it et.~al.} have performed
the marginalisation of the  $\chi^2$ function of Eq.\ (\ref{eq:chisqsn}),
over the parameters of the theory to obtain best-fit values of the parameters.
The estimated value of the quantity $E(z) = H(z)/H_0$ at the
best-fit has been shown in  left panel of Fig. 5 of \cite{Wang1},
where $H = \dot{a}/a$ is the Hubble parameter at redshift $z$ and
$H_0$ is its value at the present epoch ($z=0$). 
We have used the  results of analysis obtained by Wang {\it et.~al.}  in \cite{Wang1} along with the relations
$H=\dot{a}/a$ and $1/a = 1+z$ (value of the FRW scale factor at
present epoch is normalised to unity) and write 
$dt =  - \frac{dz}{(1+z)H_0E(z)}$
which on integration gives
\begin{eqnarray}
\frac{t(z)}{t_0} &=& 1 
- \frac{1}{H_0t_0}\int_z^0 \frac{dz^\prime}{(1+z^\prime)E(z^\prime)}
\label{eq:a2}
\end{eqnarray}
where $t_0$ is the time denoting the present epoch. 
We use the function $E(z)$ as obtained at best-fit 
from analysis of JLA data in \cite{Wang1} in Eq.\ \eqref{eq:a2}, 
and obtain $t$ as a
function of $z$ by performing the integration numerically.
We then eliminate $z$ from the obtained $z$ - $t(z)$ dependence and the
the equation $1 /a = 1 + z$ to obtain scale factor $a(t)$ as a function
of $t$. The obtained temporal profile may be used to find the time-dependence
of the Hubble parameter $H \equiv \dot{a}/a$ which directly govern the
cosmological dynamics through Friedmann equations. 
In extreme right panel of Fig.\ \ref{fig:1}, we have shown
the observed temporal behaviour of $H$ in terms of a newly introduced time parameter $\tau$, defined as 
\begin{eqnarray}
 \tau = \ln a
\label{eq:bb4}
\end{eqnarray}
Note that  present epoch corresponds to $\tau = 0$ as the scale factor at 
present epoch is normalised to unity.\\

The two independent Friedmann equations governing dynamics of late time cosmic
evolution in FRW spacetime background are 
\begin{eqnarray}
 H^2 = \frac{8\pi G}{3} (\rho_{\rm de} + \rho_{dm})\, \quad ; \quad
  \frac{\ddot{a}}{a} = - \frac{4\pi G}{3}\left[(\rho_{\rm dm} + \rho_{\rm de}) + 3p_{\rm de} \right] \,. \label{eq:bb}
\end{eqnarray}
Here we  considered a flat spacetime background (zero curvature constant)
and neglect contributions from radiation and baryonic matter during late time phase of cosmic evolution.
Using Eq.\  (\ref{eq:bb}), the 
equation of state of total dark fluid may  
be expressed  as
\begin{eqnarray}
\omega 
\equiv \frac{p_{\rm de}}{\rho_{\rm de} + \rho_{\rm dm}}
& = & -\frac{2}{3}\frac{\ddot{a}a}{\dot{a}^2} - \frac{1}{3}\nonumber\\
&=& -\frac{2}{3} \frac{a(  a^\prime HH^\prime + H^2 a^{\prime\prime})}{(a^\prime H)^2} - \frac{1}{3}  
\label{eq:bb3}
\end{eqnarray}
where in the last expression, time variable has been changed from $t$ to $\tau$ and
$^\prime$ corresponds to derivatives with respect to $\tau$.
We use the time dependence of 
the scale factor $a(\tau)$ extracted from the analysis of JLA
data to obtain the  temporal behaviour of the equation of state $\omega (\tau)$ 
of the dark fluid. The obtained time-dependence has been shown in the 
extreme left panel of Fig.\  (\ref{fig:1}).\\

Transforming the time variable from $t$ to $\tau$ and using $\omega = \frac{p_{\rm de}}{\rho_{\rm de} + \rho_{\rm dm}}$
in the   total continuity equation (Eq.\  (\ref{eq:cont1})) of the dark fluid  
we obtain
\begin{eqnarray}
\frac{d}{d\tau} \ln (\rho_{\rm de} + \rho_{\rm dm}) &=& -3 (1 + \omega(\tau)) \,,
\label{eq:bb5}
\end{eqnarray}
and performing
integration over $\tau$ we obtain 
\begin{eqnarray}
\rho(\tau)  & \equiv & \Big{[}\rho_{\rm de} + \rho_{\rm dm} \Big{]}_\tau
=
{\Big{[}\rho_{\rm de} + \rho_{\rm dm} \Big{]}_0}
\exp\left[-3\int^{\tau}_{\tau^\prime=0}(1+\omega(\tau^\prime)) 
d\tau^\prime \right] 
\label{eq:bb6}
\end{eqnarray}
We use the obtained $\tau$-dependence of the function $\omega(\tau)$ as extracted from SNe Ia data, 
in Eq.\  (\ref{eq:bb6}) to obtain temporal behaviour of the total density ($\rho_{\rm de} + \rho_{\rm dm}$) of the dark fluid. 
We have shown this dependence in the middle panel of Fig.\ \ (\ref{fig:1}). \\

As already  mentioned in Sec.\ \ref{sec:intro}, 
we assume the quantity $Q(t)$ in Eq.\  (\ref{eq:cont1}), which is a measure of rate of energy transfer between the
fluid dark matter and dark energy, to be time-dependent and
parametrise the time dependence as 
\begin{eqnarray}
Q(t) &=& Q_0 \Big{[} a(t)\Big{]}^k
\label{eq:cc1}
\end{eqnarray}
where $k$ is a constant. In terms of the  time parameter $\tau$,
the above equation reads $Q(\tau) = Q_0 e^{k\tau}$ and Eq.\  (\ref{eq:cont1}) 
takes the following form, 
\begin{eqnarray}
\frac{d{\rho}_{\rm dm}}{d\tau} + 3  \rho_{\rm dm}
 = Q_0\frac{e^{k\tau}}{H(\tau)}   
\label{eq:cc2}
\end{eqnarray}
We assume a trial solution of Eq.\  (\ref{eq:cc2}) for $\rho_{\rm dm}(\tau)$
as 
\begin{eqnarray}
\rho_{\rm dm} (\tau)& = & \Big{[}\rho_{\rm de} + \rho_{\rm dm} \Big{]}_0 \sum_{i=0}^\infty\alpha_{i}\tau^i 
\label{eq:cc3}
\end{eqnarray}
and express the temporal behaviour of $H(\tau)$
as extracted from JLA data  
in the form of  polynomial of the form
\begin{eqnarray}
\frac{1}{ H(\tau)} & = &  \sum_{i} \gamma_{i}\tau^i 
\label{eq:cc4}
\end{eqnarray}
where, the coefficients $\gamma_i$'s are obtained by fitting the above polynomial 
with the temporal behaviour of reciprocal of the function $H(\tau)$ shown
in the right panel of Fig.\ (\ref{fig:1}). The obtained best-fit values of the 
parameters $\gamma_i$'s are given in Tab.\ \ref{tab:1}.
\begin{table}[!h]
\begin{center}
\begin{tabular}{|cc|cc|c|}
\hline
$\gamma_0 = $ & 1.00118     &   $\gamma_2 =$ & -0.321574  &$\gamma_i = 0$\\
\cline{1-4}
$\gamma_1 = $ & 0.449751   & $\gamma_3 =$  & -0.149526  & for $i>3$ \\ 
\hline
\end{tabular}
\end{center}
\caption{\label{tab:1} Values of  $\gamma_i$'s in Eq.\  \eqref{eq:cc4} 
providing best fit to the values of $\frac{1}{ H(\tau)}$
extracted from analysis of  JLA data}
\end{table}
We use the Eqs.\ (\ref{eq:cc3}) and (\ref{eq:cc4}) in
Eq.\  (\ref{eq:cc2}) to  obtain
\begin{eqnarray}
\sum_{i=0}^\infty i\alpha_{i}\tau^{i-1} + 3 \sum_{i=0}^\infty\alpha_{i}\tau^i = \beta_0 \sum_{i=0}^3 \gamma_{i}\tau^i  \sum_{j=0}^\infty  \frac{k^{j}\tau^{j}}{j!}\,.
\label{eq:cc5}
\end{eqnarray}
where,
\begin{eqnarray}
\beta_0 \equiv Q_0 / [\rho_{\rm de} + \rho_{\rm dm} ]_0
\label{eq:cc6}
\end{eqnarray}
is the ratio of value of $Q$ to that of
total dark fluid energy density at present epoch.
Equating the coefficients of $\tau^{i}$ from both sides of 
Eq.\ (\ref{eq:cc5}) we obtain  
\begin{eqnarray}
\alpha_{i+1} &=&  \beta_0 \sum_{n=0}^i \frac{\gamma_{n}k^{i-n}}{(i-n)!(i+1)} - \frac{3\alpha_i}{i+1} 
\label{eq:b3}
\end{eqnarray}

We note from  Eq.\  (\ref{eq:cc3}) that 
\begin{eqnarray}
\alpha_0 &=& \frac{\rho^0_{\rm dm}}{[\rho_{\rm de} + \rho_{\rm dm} ]_0}
\label{eq:b4}
\end{eqnarray}
is the fraction of dark matter energy density contribution
to the total dark fluid density at present epoch which has the mathematically 
allowed domain as $0 \leqslant \alpha_0 \leqslant 1$. 
However, WMAP \cite{Hinshaw:2012aka} and Planck \cite{Ade:2013zuv} measurements  established that, the
observed 
value of the fraction $\alpha_0$ is $\sim 0.27$. 
For given values of $\alpha_0$, $k$, $\beta_0$  
one may find $\alpha_i$'s ($i > 0$) using Eq.\  (\ref{eq:b3}). 
The series \{$\alpha_i$\} will always  converge as 
$\gamma_i$'s are zero for $i>3$ (see Tab.\ \ref{tab:1}) and $(i+1)$ appears 
in the denominator of the recursion relation (Eq.\  (\ref{eq:b3})).
Using the evaluated series \{$\alpha_i$\}, we use Eq.\ (\ref{eq:cc3}) to
obtain the values of $\rho_{\rm dm}(\tau)$ at any any $\tau$,
corresponding to any given set of values of ($\alpha_0$, $k$, $\beta_0$).
Since, $|\tau|<1$ and the series \{$\alpha_i$\} converges,
the numerical value of $\rho_{\rm dm}(\tau)$  will have negligible contribution
from the terms above certain order in the summation series of Eq.\  (\ref{eq:cc3}).\\

\begin{figure}[t]
\begin{center}
\includegraphics[scale=0.35]{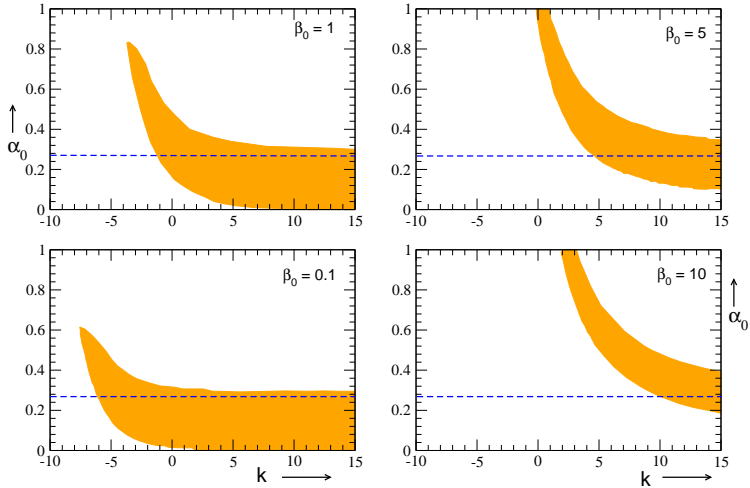} 
\end{center}
\caption{\label{fig:2}
Allowed regions in $\alpha_0 - k$ parameter space for different
benchmark values of $\beta_0$. }
\end{figure}

\begin{figure}[t]
\begin{center}
\includegraphics[scale=0.35]{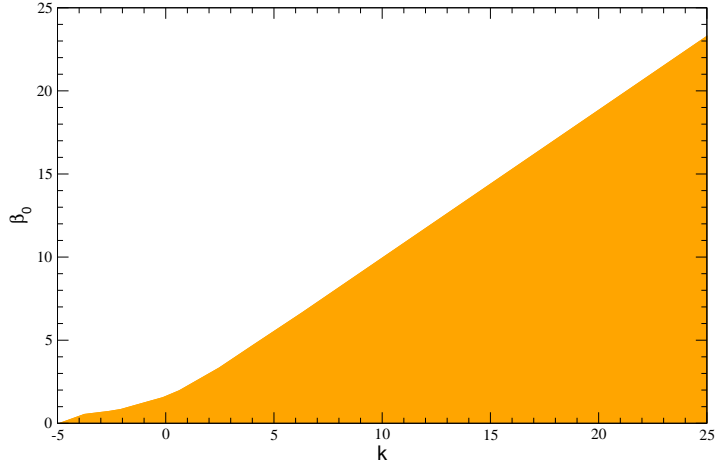} 
\end{center}
\caption{\label{fig:3}
Allowed region in $\beta_0 - k$ parameter space for $\alpha_0 = 0.27$.}
\end{figure}

The temporal behaviour of
energy density  $\Big{[}\rho_{\rm dm} + \rho_{\rm de}\Big{]}_\tau$ 
of the total dark fluid has already been obtained directly
from the analysis of JLA data (middle panel of Fig.\ \ref{fig:1}). The estimated
value of the dark matter density $\rho_{\rm dm}(\tau; \alpha_0,k,\beta_0)$
computed from Eq.\  (\ref{eq:cc3}) for given sets of values of ($\alpha_0$, $k$, $\beta_0$)
is subject to the constraint
\begin{eqnarray}
0 <  \rho_{\rm dm}(\tau;\ \alpha_0,k,\beta_0) < \Big{[}\rho_{\rm dm} + \rho_{\rm de}\Big{]}_\tau
\label{eq:c1}
\end{eqnarray}
The above constraints put limits on the range of allowed values of $\alpha_0$, $k$, $\beta_0$. 
The range of values of the parameters ($\alpha_0$, $k$, $\beta_0$) for which the condition
in  Eq.\  (\ref{eq:c1}) is satisfied, corresponds to the values of the parameters
for which the scenario of interacting DE-DM  with time-dependent diffusion coefficient
may by accommodated with the luminosity distance - redshift measurements of the SNe Ia
events in the JLA data. The constraints on parameters  ($\alpha_0$, $k$, $\beta_0$), thus 
obtained, are presented in Figs.\ \ref{fig:2} and  \ref{fig:3}. 
In Fig.\ \ref{fig:2} we have shown the allowed
domain of $\alpha_0 - k$ parameter space for 4 different 
benchmark values of the parameter $\beta_0$ (\textit{viz.} 0.1, 1.0, 5.0, 10.0).
In Fig.\ \ref{fig:3}, the allowed domain in $\beta_0 - k$ parameter space 
is shown for  $\alpha_0 = 0.27$.

\section{Realisation in terms of $k$-essence cosmology}
\label{sec:k-essence}
We also try to realise the diffusive interaction 
between DE and DM with time-dependent diffusion coefficient, in terms  
of a   $k-$essence scalar field $\phi$ representing the 
dynamics of dark energy. The scalar field $\phi$ plays the 
role of background medium in which diffusion takes place. 

\subsection{Theoretical framework of $k$-essence model}
 The minimally coupled(with gravitational field $g_{\mu\nu}$) action for a $k-$essence scalar field $\phi$ is written as,
\begin{eqnarray}
S_k &=& \int d^4x \sqrt{-g}{\cal L}(X,\phi)
\label{keq:kaction}
\end{eqnarray}
where kinetic term $X \equiv \frac{1}{2} g^{\mu\nu}\nabla_\mu\phi\nabla_\nu\phi$,
$g$ is determinant of the metric
$g_{\mu\nu}$ and $\nabla_\mu$ represents
covariant derivative associated with metric $g_{\mu\nu}$.
The total action for gravitational field $g_{\mu\nu}$ and  $k-$essence scalar field
is express as,
\begin{eqnarray}
S_k &=&  \int d^4x \sqrt{-g}\left[-\frac{R}{16\pi G_N}   + {\cal L}(X,\phi)\right]
\label{keq:actiontotal}
\end{eqnarray}
where $G_N$ is Newton's gravitational constant. We obtain the expression for
energy momentum tensor for the $k-$essence field by varying the action with respect to the field $g_{\mu\nu}$:
\begin{eqnarray}
T_{\mu\nu} & \equiv & \frac{2}{\sqrt{-g}} \frac{\delta S}{\delta g^{\mu\nu}} \nonumber\\
&=& \frac{\partial{\cal L}}{\partial X} \nabla_\mu\phi \nabla_\nu\phi -g_{\mu\nu}{\cal L}
\label{keq:emtensor}
\end{eqnarray}
the energy-momentum tensor Eq.\ (\ref{keq:emtensor}) may be written
in the form of that of a perfect fluid as
\begin{eqnarray}
T_{\mu\nu} & = & (\rho + p) u_\mu u_\nu -pg_{\mu\nu}
\end{eqnarray}
where $u_\mu$ is the effective  four-velocity  given by
\begin{eqnarray}
u_\mu &=& {\rm sgn}(\partial_0 \phi) \frac{\nabla_\mu \phi}{\sqrt{2X}}\,,
\end{eqnarray}
the pressure $p$ of the fluid is the Lagrangian density,
\begin{eqnarray}
p &=& {\cal L}(\phi,X)
\label{keq:genp}
\end{eqnarray}
and the energy density is
\begin{eqnarray}
\rho &=& 2X \frac{\partial p}{\partial X} - p
\label{keq:genrho}
\end{eqnarray}
We use the form of $k$-essence model Lagrangian as 
\begin{eqnarray}
{\cal L}(\phi,X) = V(\phi) F(X)
\label{keq:1}
\end{eqnarray}
For the $k$-essence Lagrangian of 
the form ${\cal L}(\phi,X) = V(\phi) F(X)$,
the pressure and energy density of the fluid (whose energy momentum
is equivalent to that of the $k$-essence field) may be written
from Eqs.\  (\ref{keq:genp}) and (\ref{keq:genrho}) as
\begin{eqnarray}
p &=& V(\phi) F(X) \label{keq:12} \\
\rho &=& V(\phi) (2XF_X-F)\label{keq:13} 
\end{eqnarray}
where $F_X = dF/dX$.
In this way we can relate the dark energy fluid with $k$-essence scalar field $\phi$.

\subsection{Diffusion in the background of $k$-essence scalar field}

We assume the field $\phi$
to be spatially homogeneous ($\phi = \phi(t)$) 
and the potential $V(\phi)$ to be constant ($V$). 
These respectively imply 
%
\begin{eqnarray}
X = (1/2)\dot{\phi}^2 
\label{eq:xphi}
\end{eqnarray}
and existence of a scaling 
relation of the form 
\begin{eqnarray}
XF_X^2 = C a^{-6}\,, \quad \mbox{where $C$ is a constant}
\label{eq:scaling}
\end{eqnarray}
Using the time-dependences of various  cosmological parameters extracted from SNe Ia data
as described in Sec.\ \ref{sec:observation} we obtained the 
constraints on the temporal behaviour of the field $\phi$
and the form of the function $F(X)$ for different modes
of time dependences of the diffusion coefficient, characterised in
terms of the parameter $k$. The methodology of obtaining such results 
are described below. \\

 \begin{figure}[t]
\begin{center}
\includegraphics[scale=0.35]{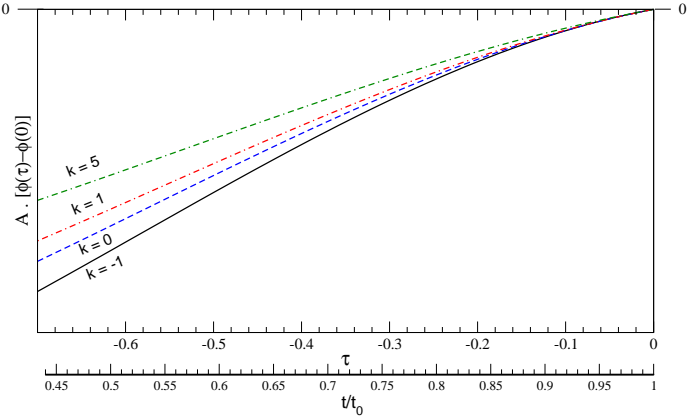} 
\end{center}
\caption{\label{fig:4}
Variation of the scalar field $\phi$ with time parameter $\tau$
and $t/t_0$. }
\end{figure}

\begin{figure}[!h]
\begin{center}
\includegraphics[scale=0.3, angle=270]{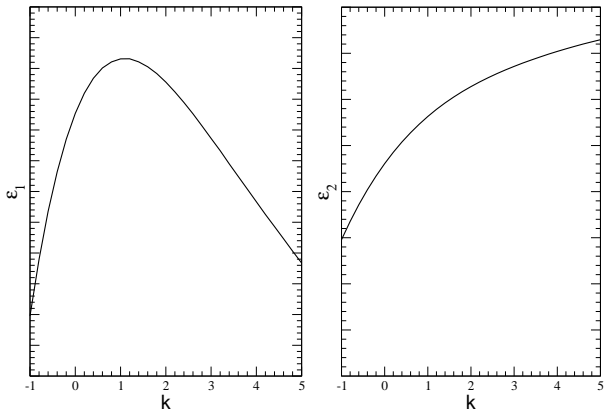} 
\end{center}
\caption{\label{fig:5}
Variation of the coefficients $\varepsilon_1, \varepsilon_2$ (in arbitrary units) with $k$.  }
\end{figure}

Using Eqs.\  (\ref{keq:12}) , (\ref{keq:13}) and (\ref{eq:scaling})  we can write,
\begin{eqnarray}
X &=& \frac{a^6 (\rho_{\rm de} + p_{\rm de})^2}{4CV^2} 
\label{eq:dk6}
\end{eqnarray}
Using Eq.\  (\ref{eq:xphi}) in Eq.\  (\ref{eq:dk6}) and 
changing the time parameter from $t$ to $\tau$ (as given in Eq.\  (\ref{eq:bb4})) we obtain
\begin{eqnarray} 
\left[\frac{\sqrt{2C}V}{(\rho_{\rm dm}^0 + \rho_{\rm de}^0)} \right] \left(\frac{d\phi}{d\tau}\right)  
&=&
\frac{a^3}{H} 
\left[\frac{\rho_{\rm de}}{(\rho_{\rm dm}^0 + \rho_{\rm de}^0)} + \frac{p_{\rm de}}{(\rho_{\rm dm}^0 + \rho_{\rm de}^0)}\right] 
\label{eq:dk8}
\end{eqnarray}
which  on Integration gives
\begin{eqnarray}
\left[\frac{\sqrt{2C}V}{(\rho_{\rm dm}^0 + \rho_{\rm de}^0)} \right] 
(\phi(\tau) - \phi_0 )
&=&
\int_{\tau^\prime=0}^\tau d\tau^\prime \left[\frac{a^3(\tau^\prime)}{H(\tau^\prime)} 
\left(\frac{\rho_{\rm de}(\tau^\prime)}{\rho_{\rm dm}^0 + \rho_{\rm de}^0} 
+ \frac{p_{\rm de}(\tau^\prime)}{\rho_{\rm dm}^0 + \rho_{\rm de}^0}\right)  \right]
\label{eq:dk9}
\end{eqnarray}
The right hand side of Eq.\  (\ref{eq:dk9}) contains the quantities
$a$ and $H$ whose observed $\tau-$dependences has already been obtained
from the analysis of JLA data as discussed in Sec.\ \ref{sec:observation}.
Since, $\omega = p_{\rm de} \Big{/} (\rho_{\rm dm} + \rho_{\rm de} )$, the last term in the right hand side of the
above equation, $p_{\rm de}(\tau)\Big{/}(\rho_{\rm dm}^0 + \rho_{\rm de}^0)$ may be expressed as 
$\omega(\tau)\rho(\tau)\Big{/}(\rho_{\rm dm}^0 + \rho_{\rm de}^0)$. This $\tau-$dependence
is known from observation as the dependences $\omega(\tau)$ and 
$\rho(\tau) \Big{/}(\rho_{\rm dm}^0 + \rho_{\rm de}^0) $
are separately known from observation as shown earlier (middle panel of Fig.\ \ref{fig:1}). 
To evaluate the other remaining term of Eq.\  (\ref{eq:dk9}),  
$\rho_{\rm de}(\tau)\Big{/}(\rho_{\rm dm}^0 + \rho_{\rm de}^0)$,
we may compute dark matter density 
$\rho_{\rm dm}(\tau; \alpha_0,k,\beta_0) \Big{/}(\rho_{\rm dm}^0 + \rho_{\rm de}^0) $ 
from Eq.\  (\ref{eq:cc3}),
corresponding to a set of values of parameters ($\alpha_0,k,\beta_0$) 
within their corresponding domains allowed from JLA data as  depicted in Fig.\ \ref{fig:2}. 
The dark energy density $\rho_{\rm de}$ may
also be evaluated at a given  ($\alpha_0,k,\beta_0$) value as
\begin{eqnarray}
\frac{\rho_{\rm de}(\tau; \alpha_0,k,\beta_0)}{(\rho_{\rm dm}^0 + \rho_{\rm de}^0)} 
&=& \frac{\Big{[}\rho_{\rm dm} + \rho_{\rm de}\Big{]}_\tau}{(\rho_{\rm dm}^0 + \rho_{\rm de}^0)} 
- \frac{\rho_{\rm dm}(\tau; \alpha_0,k,\beta_0)}{(\rho_{\rm dm}^0 + \rho_{\rm de}^0)}  \label{eq:dk10}
\end{eqnarray}
The $\tau$-dependence of the $k-$essence field, thus computed performing the integration in  Eq.\  (\ref{eq:dk9}),
would be dependent on the parameter values $\alpha_0$, $k$ and $\beta_0$. To depict the
temporal behaviour of the scalar field $\phi$ we have set $\beta_0 = 1$ and fix $\alpha_0$
at its experimental observed value at 0.27 (see discussion after Eq.\  (\ref{eq:b4})). The obtained
$\tau$ dependence of the scalar field $\phi$ for different benchmark values of the 
 parameter $k$ are shown in Fig.\ \ref{fig:4}.
We have shown the time dependence in terms of both the time parameters $\tau = \ln a(t)$ and $t$.
We find that, for any value of the diffusion parameter $k$,
the time dependence of the $k-$essence scalar field $\phi$
may be fitted in terms of polynomial of degree 2 as
\begin{eqnarray}
\phi(t/t_0, k) - \phi_0(k) &=& \varepsilon_1(k)\, (t/t_0 - 1) + \varepsilon_2(k)\, (t/t_0 -1)^2
\label{eq:dk11}
\end{eqnarray} 
where, $\phi_0(k)$ is the value of the field at present epoch ($t=t_0$),
and the coefficients 
$\varepsilon_1(k)$ and $\varepsilon_2(k)$   are functions of $k$. 
We have also computed the values of these coefficients at 
different values of the parameter $k$ and   
$k-$dependences of the 
coefficients $\varepsilon_{1,2}$ are illustrated in Fig.\ \ref{fig:5}.\\

\begin{figure}[t]
\begin{center}
\includegraphics[scale=0.35]{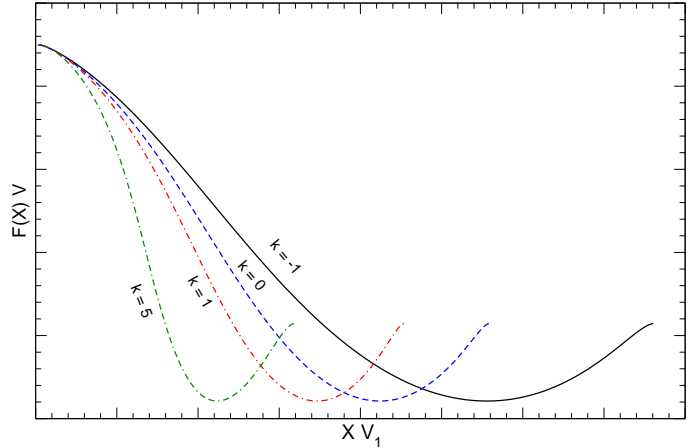} 
\end{center}
\caption{\label{fig:6}
Variation of reconstructed $F(X)V$ with $XV_1$ for different values of $k$. $V$ and $V_1$ are constants (See text for details).}
\end{figure}

Finally, in the context of realisation of time dependent diffusive DE-DM interactions
 in terms of dynamics of a $k-$essence field, we have also find
 the form of the function $F(X)$ appearing in the $k-$essence
 Lagrangian. We mentioned below how this is done using the 
 inputs from the observational data. Using Eq.\  (\ref{keq:12}) , (\ref{keq:13}) and 
 the equation $\omega(\tau) = p_{\rm de}(\tau)/\rho(\tau)$ we may write  
\begin{eqnarray}
F(X)V &=& \omega(\tau) \frac{ \rho(\tau)}{(\rho_{\rm de}^0 + \rho_{\rm dm}^0)}
\label{eq:fxv}
\end{eqnarray}
and   Eq.\ \eqref{eq:dk6} may be rewritten in the form
\begin{eqnarray}
X V_1&=& a^6 \left[\frac{\rho_{\rm de}(\tau; \alpha_0,\beta_0, k)}{\rho_{\rm de}^0 + \rho_{\rm dm}^0} 
+ \frac{\omega(\tau) \rho(\tau)}{\rho_{\rm de}^0 + \rho_{\rm dm}^0}
\right]^2
\label{eq:xv1}
\end{eqnarray}
where $V_1 = \frac{4CV^2}{(\rho_{\rm de}^0 + \rho_{\rm dm}^0)^2} $ is a constant.
Using  Eq.\  (\ref{eq:dk10}) and observed $\tau$ dependences of $a(\tau)$,  $\omega(\tau)$, $\rho(\tau)$  as
extracted from JLA data in Eqs.\  (\ref{eq:fxv}) and  (\ref{eq:xv1},) we compute
values of both the quantities $F(X)V$ and $X V_1$ 
at different values of  $\tau$ corresponding to any given set of
values of   $(\alpha_0,\beta_0, k)$. We then eliminate $\tau$ from the 
Eqs.\ (\ref{eq:fxv}) and (\ref{eq:xv1}) to obtain dependence
of $F(X) V$ as a function of $X V_1$. Thus form of the function
$F(X)$ has been extracted  
from observational data
upto the undetermined constants $V$ and $V_1$. We have shown the  variation of $F(X)$ with  $X$ in  fig.\ \ref{fig:6}
for different values of $k$ evaluated at $\beta_0=1$ and $\alpha_0 = 0.27$.

\begin{figure}[h]
\begin{center}
\includegraphics[scale=0.35]{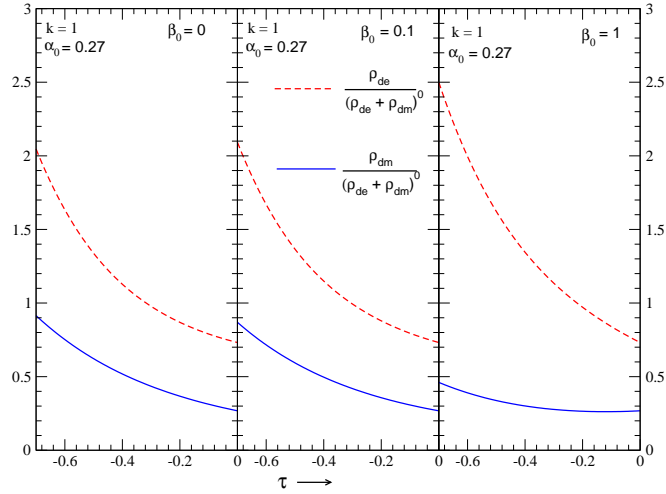} 
\end{center}
\caption{\label{fig:7}
Temporal behaviour of the energy density of dark matter and dark energy for the benchmark values of diffusion parameters ($k, \alpha_0, \beta_0$).}
\end{figure}

Figure. \ref{fig:7} depicts the diffusive activity between the dark sectors.  The energy has been transferred from dark matter to dark energy for the chosen benchmark values of the diffusion parameter $\beta_0$ (for fixed $k$ and $\alpha_0$). For $\beta_0=0$,  the continuity equation of dark matter gives $\rho_{\rm dm} \propto \frac{1}{a^3}$ which corresponds to the non-interacting scenario between these two sectors. We also choose two non-zero (positive) values of  $\beta_0$ to realise  the interacting nature of this model.  The figure shows for the larger values of $\beta_0$ corresponds to the higher amount of transferred energy from dark matter to dark energy. The figure also represents dark matter density is always lower than dark energy density for the higher values of $\beta_0$ at any epoch.

\section{Conclusion}

In this work, we study the  scenario of interacting DE and DM, with a time-dependent 
diffusive interaction between them. We have shown that such a model can be accommodated
within the observations of luminosity distance - redshift data  in SNe Ia events.
We obtain the constrains on  different relevant parameters of this model from the observational data. 
The two parameters of the model which are relevant in the context of this work
are $Q_0$ and $k$. They parametrize the rate of exchange of energy due to
diffusion from dark matter to dark energy in a form  
$Q_0 [a(t)]^k$ appearing in the non-conservation equation\  (\ref{eq:cont1}).
For convenience, in stead of working with the parameter $Q_0$,
we have chosen $Q_0/(\rho_{\rm de}^0 +\rho_{\rm dm}^0 ) \equiv \beta_0$ as the parameter.
Besides $\beta_0$ and $k$, the   parameter
$\alpha_0 \equiv \rho_{\rm dm}^0/(\rho_{\rm de}^0 +\rho_{\rm dm}^0 )$,
which is approximately the fractional of contribution of dark matter dark density
to the total density at present epoch also appears in the framework of 
our analysis.  \\

We have   shown the temporal behaviour of the Hubble parameter, $H$, total equation 
of state ($\omega$) and total energy density $(\rho)$ of the dark fluid 
as extracted from JLA data in Fig.\ \ref{fig:1}. We have exploited these dependences 
to obtain the constraints on the above mentioned parameters $\beta_0, k, \alpha_0$. 
The obtained constraints are
depicted in
Figs.\ \ref{fig:2} and  \ref{fig:3}.  
The value of the parameter $\alpha_0$ is however, independently determined
from WMAP and PLANCK experiments as $\alpha_0 \approx 0.27$.
The results show that, if we choose this value of $\alpha_0 (\approx 0.27)$,
 the allowed values of the parameter $k$ has a lower limit.  For example,
for different values of $\beta_0$, the obtained values of the lower limit of $k$ 
are given in Tab.\ \ref{tab:2}.\\

\begin{table}[!h]
\begin{center}
\begin{tabular}{|c|c|c|c|c|}
\hline
values of $\beta_0$ & 0.1      & 1.0   &    5.0 & 10.0  \\
\hline 
lower limit of $k$ &  -6.13     &-1.11   &  4.76 &  9.89 \\ 
\hline
\end{tabular}
\end{center}
\caption{\label{tab:2}  Lower limit of allowed values of parameter $k$
for different values of $\beta_0$ with $\alpha_0 = 0.27$.}
\end{table}

In addition to this, 
in the context of the interacting DE-DM model considered here along with the
constraints on relevant parameters of the model from observational data, 
we also  address certain issues related to $k$-essence scalar field model of dark
energy. We consider a homogeneous scalar field $\phi$ driven by a $k-$essence Lagrangian
(with constant potential)
to represent dynamics of  dark energy. We assume, in  
DE-DM interacting scenario considered here, the diffusion from
dark matter takes place in the background medium of the
scalar field. Using the observational features of the cosmological
parameters as extracted from the JLA data as inputs, 
we  find constraints on the time-dependence of the field $\phi$.
The existence 
of scaling relation (Eq.\ \ref{eq:scaling}) owing to constancy of
the potential $V$ in the $k-$essence Lagrangian, also
enables us to obtain the form of the function $F(X)$  appearing in the Lagrangian.
We found that the temporal behaviour of the 
scalar field, in this context, may be 
very convincingly accommodated within a profile
$\phi(t/t_0,k) - \phi_0(k) = \varepsilon_1(k) (t/t_0-1) + \varepsilon_2(k) (t/t_0-1)^2 $,
where coefficients $\varepsilon_{1,2}$ are dependent on the (diffusion)
parameter $k$ and these dependences are also obtained using 
inputs from observation (See Fig.\ \ref{fig:5}).  
The obtain form of the function $F(X)$ for different 
values of the parameter $k$ are depicted in 
Fig.\ \ref{fig:6}.  Diffusive nature of DM and DE components have been achieved through the parametrize form of $Q(t)$, fig.\ \ref{fig:7} represents the validity of our model in both non-interacting and interacting scenarios.

\section{Acknowledgements}
The authors are grateful to the referee for very helpful suggestions.
A.C. would like to thank University Grants Commission (UGC), India
for supporting this work by means of NET Fellowship (Ref.No.22/06/2014(i)EU-V
and Sr.No.2061451168).

\end{document}